\documentclass[english]{article}
\usepackage[LGR,T1]{fontenc}
\usepackage[latin9]{inputenc}
\usepackage{textcomp}
\usepackage{amsmath}
\usepackage{amssymb}
\usepackage{cancel}
\usepackage{stackrel}

\makeatletter

\DeclareRobustCommand{\greektext}{%
  \fontencoding{LGR}\selectfont\def\encodingdefault{LGR}}
\DeclareRobustCommand{\textgreek}[1]{\leavevmode{\greektext #1}}
\ProvideTextCommand{\~}{LGR}[1]{\char126#1}

\newcommand{\lyxaddress}[1]{
	\par {\raggedright #1
	\vspace{1.4em}
	\noindent\par}
}

\makeatother

\usepackage{babel}
\begin{document}
\title{Effective Field Theory Framework: Construction Strategies and Soft
Collinear Effective Theory (SCET)}
\author{Waqas Riaz}
\maketitle

\lyxaddress{Department of Physics, Quaid-I-Azam University, Islamabad 45320}
\begin{abstract}
Effective Field Theory (EFT)\cite{key-1} stands as a cornerstone
in modern theoretical physics, offering a powerful framework for describing
the dynamics of physical systems across a wide range of energy scales.
This article provides an in-depth exploration of EFT and its diverse
applications in various branches of physics. Beginning with a foundational
overview of EFT principles, including its formulation, renormalization,
and power counting rules, we delve into its versatility in addressing
complex phenomena beyond the reach of traditional approaches. In particle
physics, EFT techniques are indispensable for precision calculations
and theoretical predictions, enabling the systematic treatment of
quantum field theories at energies inaccessible to direct experimentation.
Moreover, EFT plays a crucial role in understanding the low-energy
dynamics of hadrons, nuclei, and other strongly interacting systems.
This article aims to provide a comprehensive introduction to EFT and
as an example of constructing EFT from a fundamental theory like QCD
we explore how to construct a soft collinear effective theory SCET.
\end{abstract}
\pagenumbering{arabic}
\setcounter{page}{1}

\section{Effective Field Theory (EFT)}

An EFT, is basically a quantum theory which is a kind of approximation
for an underlying physical theory. The formalism of EFT is based on
particular desired scale. In EFTs, the physical quantities are expanded
in the small ratio of the scales to separate the low-energy contributions
from the high-energy part. We expand the original theory up-to arbitrary
precision and construct EFT from it. As we know that there are different
length scales in universe (from macroscopic to microscopic) and effective
field theory enable us to compute things merely in particular desired
scale.

There is an interesting physics at all macroscopic and microscopic
scales. On the basis of scale and domain of knowledge the physics
is divided into different branches, i.e., Quantum field theory, string
theory, general relativity, Newtonian gravity, classical mechanics
and classical electrodynamics.

\subsection{Motivation Behind Effective Field Theory (EFT)}

Most of the time we rely on an EFT description of a physical system
because of the unavailability of a more fundamental theory. The key
advantages of an EFT approach can be
\begin{itemize}
\item By focusing only on the relevant degrees of freedom, and neglect the
irrelevant ones in EFT the calculations become simple.
\item New symmetries appears through relevant degrees of freedom, that otherwise
would have remained hidden. Hence, we became aware about the conserved
quantities in our problem after detecting the symmetries.
\item EFT provide a useful and simple way of testing models of physics beyond
the SM.
\end{itemize}

\subsection{Steps involved in developing EFT}

Mainly following three steps are taken while constructing EFT:
\begin{itemize}
\item Determine relevant degrees of freedom.
\item Symmetries.
\item Power counting .
\end{itemize}

\subsubsection{Relevant degrees of freedom}

We only keep those parameters which actually matter for the problem,
i.e., in order to study low energy properties of SM we neglect the
heavy particles and only keep the lighter ones. The heavy and light
particles have their distinct degrees of freedom and on the basis
of our requirement we keep the relevant degrees of freedom and integrate
out the irrelevant ones.

\subsubsection{Symmetries}

Symmetries play a key role in constructing the EFT. By imposing the
symmetries on the Lagrangian of underlying problem, it is possible
predict the behavior of the system without solving the underlying
complex equations of motion. This makes EFT a powerful tool for studying
complex physical systems. There are continuous, discrete and global
symmetries. There may be a case when our problem has no inherent or
apparent symmetry but expanding the problem give some symmetry. These
are called the emerging symmetries. The EFT has more symmetry than
original underlying theory. 

\subsubsection{Power counting}

Power counting is a method that enable us to determine the significance
of different terms appear in the effective Lagrangian. It help us
to determine which terms in the Lagrangian are most significant. Power
is assigned to the terms in the effective Lagrangian based on the
number of fields and derivatives that they contain. Terms with a higher
power are considered to be more important, as they are expected to
have a large influence on the behavior of the system, whereas terms
with a lower power are typically considered to be less important,
and they can be neglected in calculations.

\section{Formalisms of Effective Field Theories}

In general, EFT is constructed in two ways:
\begin{itemize}
\item Top-down EFT.
\item Bottom-up EFT.
\end{itemize}

\subsection{Top-down EFT}

In this we choose an upper and lower cut-off and then subsequently,
remove the field modes that lies at momenta above the desired level.
We perform this by writing down a general Lagrangian for the light
fields in the form of a sum of all allowed operators, with unknown
coefficients. In top-down EFT, the full theory is known and we need
a simpler theory (EFT) at low energies that we construct from full
theory. We integrate out or remove heavier particles and match onto
a low energy theory which yields new operators and low energy couplings.
We can say

\[
Theory-1\longrightarrow Theory-2,
\]
where Theory-1 and Theory-2 are high and low energy theories, respectively.
As mentioned above, in top-down theory we move from higher energy
scale to lower energy scale, i.e., we remove heavier particles from
original theory and only keep the lighter ones to construct the EFT.
Mathematically, it is expressed as

\[
\mathcal{L}_{High}\rightarrow\sum_{n}\mathcal{L}_{Low}^{(n)}.
\]
We expand the Lagrangian up to an arbitrary precision. Desired precision
tells us where to stop and this can be observed through experiment
or by some suitable guess. The experiments tell us up-to how much
$n$ we have to calculate and where to stop. $\mathcal{L}_{High}$
and $\mathcal{L}_{Low}$ agrees in infrared region but disagree in
ultraviolet (UV) region, hence they might have different UV divergences.
Following are the examples of Top-down effective field theory:
\begin{itemize}
\item Heavy Quark Effective Theory (HQET)\cite{key-2}.
\item NRQCD and QED.
\item Soft Collinear Effective Theory (SCET)\cite{key-3}.
\end{itemize}

\subsubsection{Heavy Quark Effective Theory (HQET) as top-down EFT}

HQET is a classic example of non relativistic effective theory. The
$u$, $d$ and $s$ have mass less than the $\Lambda_{QCD}$ scale.
The top quark do not make a bound state because it is heavy and have
short life-time. The $c$ and $b$ quarks are also termed as heavy
quarks because there mass is greater than $\Lambda_{QCD}.$ In a bound
state system $Q\bar{q}$, where $Q$ is the heavy quark and the interactions
with the light-quark consist of the non-perturbative dynamics of gluon
exchanges or $q\bar{q}$ pairs are of order $\Lambda_{QCD}.$ In HQET,
we do not remove the heavy quarks from the theory but we see the influence
of light degrees of freedom on the heavy quark or we can say that
we measure the fluctuations near heavy particle by tickling it with
light particle. In quark model the degrees of freedom are $Q\bar{q}$
mesons. We deal with mass scale and size in this problem, i.e., 

\[
r^{-1}\backsim\land_{QCD}\ll m_{Q}.
\]
The fluctuations of $Q$ due to light degrees of freedom is given
by

\begin{equation}
\underset{m_{Q}\rightarrow\infty}{\text{lim}}\mathcal{L}=\underset{m_{Q}\rightarrow\infty}{\text{lim}}\bar{Q}\left(i\cancel{D}-m_{Q}\right)Q,
\end{equation}
where $D_{\mu}=\partial_{\mu}-igA_{\mu}^{a}T^{a}$ and $m_{Q}\rightarrow\infty$
is low energy limit of QCD scale.

Consider a propagator for heavy quark. It has to be on-shell up, so
we can write

\[
p^{\mu}=m_{Q}v^{\mu}+k^{\mu},
\]
where $k^{\mu}\thicksim\land_{QCD}$ shows how off-shell the heavy
quark is. Writing

\begin{align*}
i\frac{\left(\cancel{p}+m_{Q}\right)}{p^{2}-m_{Q}^{2}+i\epsilon} & =\frac{i\left(m_{Q}\cancel{v}+m_{Q}+\cancel{k}\right)}{2m_{Q}v.k+k^{2}+i\epsilon}\\
 & =i\left(\frac{1+\cancel{v}}{2}\frac{1}{v.k+i\epsilon}+\mathcal{O}\frac{1}{m_{Q}}\right).
\end{align*}
If we deal with an $n$ point function, there will be propagator around
the vertex, therefore,

\begin{align*}
\frac{1+\cancel{v}}{2}\gamma^{\mu}\frac{1+\cancel{v}}{2} & =\frac{1}{4}\left[\gamma^{\mu}+\left(\cancel{v},\gamma^{\mu}\right)+\cancel{v}\gamma^{\mu}\cancel{v}\right]\\
 & =\frac{1}{4}\left[2v^{\mu}+\cancel{v}2v^{\mu}\right]\\
 & =\frac{1+\cancel{v}}{2}v^{\mu}\frac{1+\cancel{v}}{2}.
\end{align*}
The interaction is spin-independent so HQET Lagrangian is

\begin{equation}
\mathcal{L}_{HQET}=\bar{Q}_{v}iv.DQ_{v},
\end{equation}
where $\text{\ensuremath{Q_{v}}}$ field satisfies $\frac{1+\cancel{v}}{2}Q_{v}=\text{\ensuremath{Q_{v}} }$and
$iv.D=iv.\partial-gv.A,$ so this field describe heavy quark. Now
let us start with a convenient field redefinition

\[
Q(x)=e^{-im_{Q}v.x}\left[Q_{v}(x)+B_{v}(x)\right].
\]
The projection relations of fields $Q_{v}$ and $B_{v}$ are

\begin{align*}
Q_{v} & =\text{\ensuremath{e^{im_{Q}v.x}}}\frac{1+\cancel{v}}{2}Q(x),\\
B_{v} & =\text{\ensuremath{e^{im_{Q}v.x}}}\frac{1-\cancel{v}}{2}Q(x).
\end{align*}
where $e^{im_{Q}v.x}$ is a phase factor. 

The momentum of $Q_{v}$ and $B_{v}$ is the same as that of $Q$
and shifted by $m_{Q}v_{\mu}$ as shown below

\begin{equation}
\mathcal{P}_{\mu}Q_{v}(x)=\text{\ensuremath{e^{im_{Q}v.x}}\ensuremath{\frac{1+\cancel{v}}{2}}}\left(m_{Q}v_{\mu}+\mathcal{P}_{\mu}\right)Q(x).
\end{equation}
$\text{\ensuremath{Q_{v}}}$ field is massless while $\text{\ensuremath{B_{v}}}$
field is very heavy, i.e.,
\[
\text{\ensuremath{Q_{v}=\frac{1+\cancel{v}}{2}Q_{v},} }
\]
 and
\[
\text{\ensuremath{B_{v}=\frac{1-\cancel{v}}{2}B_{v}.} }
\]
We can break the $\cancel{D}$ into two pieces as

\[
i\cancel{D}=\cancel{v}iv.D+i\cancel{D}_{T},
\]
where $D_{T}^{\mu}=D^{\mu}-v^{\mu}v.D$ such that; $v.D_{T}=0$. Here,
$D_{T}$ is the transverse piece.

\begin{align*}
\mathcal{L}_{QCD} & =\left[Q_{v}+\bar{B_{v}}\right]e^{im_{Q}v.x}\left\{ \text{\ensuremath{\cancel{v}iv.D+i\cancel{D}_{T}}}-m_{Q}\right\} \text{\ensuremath{e^{-im_{Q}v.x}\left(Q_{v}+B_{v}\right)} }\\
 & =\text{\ensuremath{\left[\bar{Q_{v}}+\bar{B_{v}}\right]e^{im_{Q}v.x}e^{-im_{Q}v.x}}}\left(\cancel{v}-1\right)m_{Q}+\left(\cancel{v}iv.D+i\cancel{D}_{T}\right)\left(Q_{v}+B_{v}\right).
\end{align*}
Here, we have used, $\cancel{D}_{T}\frac{1-\cancel{v}}{2}=\frac{1+\cancel{v}}{2}\cancel{D}_{T}$.

As it is mentioned above that $Q_{v}$ field is massless while $B_{v}$
field is heavy, due to this $\text{\ensuremath{B_{v}}}$ decouples
in limit of $m_{Q}\rightarrow\infty$. If we consider only external
fields then as $m_{Q}\rightarrow\infty$, $B_{v}$ decouples. Hence

\begin{equation}
\mathcal{L}_{HQET}=\bar{Q_{v}}iv.DQ_{v}.
\end{equation}
Physically, $Q_{v}$ corresponds to heavy particle and $B_{v}$ corresponds
to heavy anti-particles. If phase redefinition is in-phase then we
expand about the particles and similarly if we choose the opposite
phase then we expand around anti-particles. Following are important
points that are necessary to be mentioned here:
\begin{itemize}
\item The above field redefinition is at tree level and valid up-to leading
order in $\frac{1}{m_{Q}}$ and $\alpha_{s}\left(m_{Q}\right)$. It
well described the couplings of quarks to gluons at leading orders.
\item The anti-particles are integrated out, and it is easiest to see that
if we go to rest frame 
\end{itemize}
\[
v_{r}^{\mu}=\left(1,0,0,0\right){}^{T},
\]
which gives $\frac{1+\cancel{v}}{2}=\frac{1+\gamma_{0}}{2}$.

The gamma matrices in Dirac's representation are 

\begin{align*}
\gamma^{0} & =\left(\begin{array}{c}
1\\
0
\end{array}\begin{array}{c}
0\\
-1
\end{array}\right),\\
\gamma^{i} & =\left(\begin{array}{c}
0\\
-\sigma^{i}
\end{array}\begin{array}{c}
\sigma^{i}\\
0
\end{array}\right).
\end{align*}
where $\sigma^{i}$ represents the 2x2 Pauli's matrices. 

The Dirac's representation clearly shows the difference between particles
and antiparticles

\[
\frac{1+\gamma^{0}}{2}u_{Dirac}=\left(\begin{array}{c}
\psi_{v}\\
0
\end{array}\right),
\]
where $u_{Dirac}$ is a Dirac's spinor. 

\subsubsection{Bottom-up EFT}

In bottom-up EFT, the underlying theory is unknown or there is a possibility
that high energy theory is known but doing matching calculations to
integrate out degrees of freedom for those calculations could be very
difficult. Without doing calculations in high energy theory in order
to carried out low energy theory is difficult. In this case

\[
Theory-2\longrightarrow Theory-1,
\]
where Theory-1 and Theory-2 are high and low energy theory respectively.

Following are the examples of bottom-up EFT:
\begin{itemize}
\item Chiral perturbation theory (ChPT),
\item Standard model effective field theory (SMEFT).
\end{itemize}

\subsubsection{SMEFT as a bottom-up EFT}

SM is not an ultimate theory of nature and that is why it is considered
as a classic example of a bottom-up EFT. Generally, the Lagrangian
for bottom-up EFT is constructed as 

\begin{equation}
\sum^{(n)}\mathcal{L}_{Low}^{(n)}=\mathcal{L}^{(0)}+\mathcal{L}^{(1)}+\mathcal{L}^{(2)}+\cdots,
\end{equation}
where zeroth order $\mathcal{L}^{(0)}$ is the SM Lagrangian.

In SM there are lots of different scales and from top-down EFT point
of view, there is a need to get rid of top quark, $W$ and $Z$ bosons,
Higgs boson and then proceed further. The lowest order Lagrangian
of SM comprises of Lagrangian of gauge bosons, fermions, Higgs bosons
and right handed neutrinos, i.e.,

\begin{equation}
\mathcal{L}^{(0)}=\mathcal{L}_{Gauge}+\mathcal{L}_{Fermions}+\mathcal{L}_{Higgs}+\mathcal{L}_{NR}.
\end{equation}
The gauge bosons Lagrangian is 

\begin{equation}
\mathcal{L}_{Gauge}=-\frac{1}{4}B^{\mu\nu}B_{\mu\nu}-\frac{1}{4}W_{\mu\nu}^{a}W_{a}^{\mu\nu}-\frac{1}{4}G_{\mu\nu}^{A}G_{A}^{\mu\nu}.
\end{equation}
The fermionic Lagrangian is

\begin{equation}
\mathcal{L}_{Fermions}=\sum_{\psi_{L}}\bar{\psi}_{L}\text{\ensuremath{i\phi\psi_{L}}}+\sum_{\psi_{R}}\bar{\psi}_{R}\text{\ensuremath{i\phi\psi_{R},} }
\end{equation}
where,$\underset{\psi_{L}}{\sum}\bar{\psi}\text{\ensuremath{\iota\phi\psi_{L}}}$
and $\underset{\psi_{R}}{\sum}\bar{\psi}\text{\ensuremath{\iota\phi\psi_{R}}}$
are sum over left and right handed fields interacting with the Higgs
boson. Now, the covariant derivative can be decomposed as

\begin{equation}
iD_{\mu}=i\partial_{\mu}+g_{1}B_{\mu}^{'}+g_{2}W_{\mu}^{a}T^{a}+gA_{\mu}^{a}T^{a},
\end{equation}
where $g_{1}$ is the gauge coupling for hyper charge, $g_{2}$ is
the coupling for $SU(2)$ weak interactions, and $g$ is the gauge
coupling for QCD.

\subsubsection{Dimensional power counting in SMEFT}

The power counting in this bottom-up EFT approach of SM can be done
by expanding mass scale Epsilon ($\epsilon$). The power counting
for the SM as an EFT, is based on a new mass scale that we have ignored
or neglect at the higher energy $\lambda_{new}$ while studying low
energy properties of SM. The $\epsilon$ is basically a ratio of masses
of contents of SM to the contents that we left in our discussion.
$\epsilon=\frac{M_{(SM)}}{\lambda_{new}},$ where $M_{(SM)}$ is the
sum of quarks mass, $W$ and $Z$ bosons mass, Higgs boson mass and
all other mass scales of the SM, while $\lambda_{new}$ is the contents
that we are going to ignore in our discussion. By keeping the regard
of EFT, if we have grand unified theory (GUT) and broken supersymmetry
then it would also be a part of $\lambda_{new}$ as

\begin{equation}
\epsilon=\frac{M_{Higgs}+M_{Quarks}+M_{W}+M_{Z}+M_{Leptons}+\cdots}{M_{Planck}+M_{Grand-Unified-Theory}+M_{Supersymmetry-broken}},
\end{equation}
where $M_{Planck}$ is Planck's mass whose numerical value is approximately
equal to $2.176434$$\times$$10^{-8}kg$.

Hence, we have concluded that from EFT point of view any physics that
we have left out of SM description will be fitted in denominator.

\subsubsection{Renormalizable field theory}

A theory is renormalizable if at any order the UV divergence can be
absorbed into a finite number of parameters. The EFT point of view
about renormalizability is more general because EFT has an idea of
power counting. Consider the scalar field theory in $d$-dimensions
where the mass dimensions $m$ determines the power counting

\begin{equation}
S\left[\phi\right]=\int d^{d}x\left(\frac{1}{2}\partial_{\mu}\phi\partial^{\mu}\phi-\frac{1}{2}m^{2}\phi^{2}-\frac{\lambda}{4!}\phi^{4}-\frac{\tau}{6!}\phi^{6}\cdots\right),\label{eq:1.4.11}
\end{equation}
where $\frac{1}{2}\partial_{\mu}\phi\partial^{\mu}$ is standard kinetic
term, $\frac{1}{2}m^{2}\phi^{2}$ is standard mass term, $\frac{\lambda}{4!}\phi^{4}$
is $\phi^{4}$ scalar field theory term, $\frac{\tau}{6!}\phi^{6}$
is $\phi^{6}$ scalar field theory term.

Now dimensions of the terms involved in Eq. (\ref{eq:1.4.11}), is
determined are $[d^{d}x]=-d$, $[m^{2}]=2$, $[\lambda]=4-d$,$[\tau]=6-2d$.
Hence, we can say that $[\lambda]$ and $[\tau]$ has dimensions of
zero, if $d$ is 4 and 3, respectively. The dimensions of $[\lambda]$
and $[\tau]$ depends upon value of $d$ while that of $\left[m^{2}\right]$
is independent of $d$. The correlation function of a bunch of different
$\phi's$ at different spacetime point is

\[
\left\langle \phi(x_{1})\cdots\phi(x_{n})\right\rangle ,
\]
if we look at large distance then we can make $x$ large as 

\[
x^{\mu}=Sx'^{^{\mu}},
\]
where $x'$ is fixed and $S$ $\rightarrow\infty$.

Now let us redefine the field

\[
\phi(x)=S^{2-d/2}\phi'(x'),
\]
 $\phi(x)$ and $\phi'(x)$ is new and old field respectively. The
$\phi'$ field is 

\begin{equation}
S^{'}\left[\phi'\right]=\int d^{d}x\left(\frac{1}{2}\partial_{\mu}\phi'\partial^{\mu}\phi'-\frac{1}{2}m^{2}S^{2}\phi'^{^{2}}-\frac{\lambda}{4!}S^{4-d}\phi'^{^{4}}-\frac{\tau}{6!}S^{6-2D}\phi'^{^{6}}\right).
\end{equation}
The correlation function in term of $\phi'$ is just a function of
$x'$ and $x'$ is fixed as

\begin{equation}
\left\langle \phi\left(Sx_{1}'\right)\cdots\phi\left(Sx_{n}'\right)\right\rangle =S^{n(2-d)/2}\left[\left\langle \phi'\left(x_{1}'\right)\cdots\phi'\left(x_{n}'\right)\right\rangle \right].
\end{equation}
Now taking $S\rightarrow\infty$ and $d=4$ then, $m^{2}$ term become
more and more important, $\tau$ term become less important and $\lambda$
is equally important as it was before. On the basis of these observations
we can conclude that $\phi^{2}$ is relevant, $\phi^{4}$ is marginal
and $\phi^{6}$ is irrelevant operator. We are controlling the power
counting with some common parameter $S$. As $S$ is in our control
to look at large distances and it is connected with the dimensions
of operators i.e., $\phi^{2},\phi^{4},\phi^{6}$ where $2,4,6$ are
dimensions of operator $\phi.$ Now let us take $S$ finite but large
as we want to study a long distance behavior. Large distance simply
means small momenta, i.e., $p<<\lambda_{new}$. It is necessary to
mention that relevant terms are more important than marginal terms
and marginal terms are important than irrelevant ones. From previous
task i.e., $S\rightarrow\infty$, now we know that by looking at mass
dimensions we can do power counting. We just associate a power counting
to the parameters as $(m^{2})\approx(\:\Lambda_{new}^{2}),$$(\lambda)\approx(\Lambda_{new}^{0})$,
$(\tau)\approx(\Lambda_{new}^{-2}).$ Where $(m^{2}),(\lambda)\text{ }and$
$(\tau)$ are relevant, marginal and irrelevant operators, respectively.
Since in our case, the EFT looks toward the IR of the underlying theory,
the mass term of the heavy particle will not be included. The $\phi^{4}$
and $\phi^{6}$ terms are included and they can usually be integrated
out, leaving an EFT that contains only light degrees of freedom.

\section{Effective field theory approach for H-atom}

Let us briefly discuss the H-atom from effective field theory point
of view. While describing H-atom on large scale we do not need to
know about the quark contribution. The binding energy of H-atom is 

\begin{equation}
E_{o}=\frac{1}{2}m_{e}\alpha^{2}.
\end{equation}
Therefore, if we want to study H-atom on infinitesimally small scale
then we include the quark contribution in its binding energy. Hence,
above result will be modified as \cite{key-4}

\begin{equation}
E_{o}=\frac{1}{2}Me\alpha^{2}\left[1+\mathcal{O}\left(\frac{m_{e}^{2}}{m_{b}^{2}}\right)\right],
\end{equation}
where $\left[1+\mathcal{O}\left(\frac{m_{e}^{2}}{m_{b}^{2}}\right)\right]$
is $b$ quark contribution in binding energy and $\mathcal{O}\left(\frac{m_{e}^{2}}{m_{b}^{2}}\right)$
is an arbitrary constant. The quark contribution in binding energy
is fairly very small, i.e., up-to an order of $10^{-8}$. So, while
developing EFT for H-atom at large scale we ignore the quark contribution.
If we want to study the low energy properties of H-atom then we can
not ignore the quark contribution. The electromagnetic coupling $\alpha$
has different values. If we evaluate electromagnetic coupling at the
scale like $W$ boson mass then its value is $\alpha\left(M_{W}\right)\approx\frac{1}{128}$.
Similarly, if we evaluate at very low energy scale, i.e., the mass
of electron or below, then it is $\alpha\left(0\right)\approx\frac{1}{137.03}$.
The variation in the value of electromagnetic coupling constant arises
due to quark contribution at small scale. While solving H-atom in
ordinary quantum mechanics we do not introduce the quark contribution
because the typical momentum transfer in H-atom is $\vec{p}=m_{e}\alpha$,
which is very less than proton size. So the typical protons that are
involved in binding H-atom just have much lower energy so we do not
need to know about quarks inside the proton.

\section{Electroweak Effective Hamiltonian}

In order to deal with a problem at a specific momentum scale, it is
convenient to do calculations only with the relevant degrees of freedom.
Disentangling the irrelevant degrees of freedom makes our calculations
simple and easier. Physically, it implies that the long distance physics
should be independent of short-distance physics and vice-versa. For
example, while studying a $b$ quark decay, we should not worry about
the gravitational interactions or in studying a hydrogen atom, the
effect of $b$ quark on the vacuum polarization diagram is negligible
and one only needs the QED contribution at this scale. If we probe
a physics at a hadronic scale, which is lower than the weak interactions
scale we only need to integrate out the heavy degrees of freedom and
replace them with corresponding effective vertices. Moreover, we need
to integrate out heavy particles at once because by integrating out
the top quark only, the third generation $SU_{L}(2)$ symmetry would
break. This process of replacing heavy degrees of freedom is tantamount
to finding a weak effective Hamiltonian (WEH). This Hamiltonian can
then be used to study the $B$-decays at the desired scale without
worrying to begin from scratch. The SM symmetries still hold in the
WEH, although new symmetries may appear in the effective theory during
calculations. To carry out this procedure, we rely upon an important
tool known as the operator product expansion (OPE). The OPE is based
on the fact that the different local operators at different space-time
positions can be written as a sum over composite local operators.
Mathematically, it is written as\cite{key-5}

\begin{equation}
\underset{x\rightarrow y}{lim}O_{1}(x)O_{2}(y)=\underset{n}{\sum}C_{n}(x-y)O_{n}(x),
\end{equation}
where $C_{n}$ is the Wilson coefficient. 

The amplitude is given in unitary gauge as\cite{key-6}

\begin{equation}
\text{\ensuremath{\mathcal{M}=\left(\frac{ig}{\sqrt{2}}\right)^{2}V_{cb}V{}_{ud}^{\ast}\left(g^{\mu\nu}-\frac{k^{\mu}k^{\nu}}{M_{W}^{2}}\right)\left(\frac{1}{k^{2}-M_{W}^{2}}\right)-\text{\ensuremath{\left[\bar{c}\gamma_{\mu}\left(1-\gamma^{5}\right)b\right]\left[\bar{d}\gamma_{\nu}\left(1-\gamma^{5}\right)u\right]}}}},
\end{equation}
where $V_{cb}V{}_{ud}^{\ast}$ are the CKM matrix elements, and $k$
is the momentum transfer. As $k$ is very small as compared to $M_{W}$,
hence we can ignore the $k^{2}$ term and above equation can be modified
as 

\begin{equation}
\mathcal{M}=-\frac{4G_{F}}{\sqrt{2}}V_{cb}V{}_{ud}^{\ast}\text{\ensuremath{\left[\bar{c}\gamma_{\mu}\left(1-\gamma^{5}\right)b\right]\left[\bar{d}\gamma_{\nu}\left(1-\gamma^{5}\right)u\right]},}
\end{equation}
where $G_{F}$ is the Fermi coupling constant. The amplitude in above
equation is in terms of local operators. The weak Hamiltonian for
$B$-decays is \cite{key-7}

\[
\text{\ensuremath{H_{EWH}}=\ensuremath{-\frac{G_{F}}{\sqrt{2}}V_{ts}^{*}V_{tb}\left(\stackrel[i=1]{10}{\sum}C_{i}(\mu)Q_{i}(\mu)\right)}},
\]
where $C_{i}(\mu)$ are the Wilson coefficients at scale $\mu\thicksim m_{b}$
found by integrating out the heavy particles in the corresponding
diagrams. The operators corresponding to $Q_{i}$ are \cite{key-8}

\begin{align}
Q_{1} & =\left(\bar{s}_{L}\gamma_{\mu}T^{a}q_{L}\right)\left(\bar{q}_{L}\gamma^{\mu}T^{a}b_{L}\right),\nonumber \\
Q_{2} & =\left(\bar{s}_{L}\gamma_{\mu}q_{L}\right)\left(q_{L}\gamma^{\mu}b_{L}\right),\nonumber \\
Q_{3} & =\left(\bar{s}_{L}\gamma_{\mu}b_{L}\right)\underset{q}{\sum}\left(\bar{q}\gamma^{\mu}q\right),\nonumber \\
Q_{4} & =\left(\bar{s}_{L}\gamma_{\mu}T^{a}b_{L}\right)\underset{q}{\sum}\left(\bar{q}\gamma^{\mu}T^{a}q\right),\nonumber \\
Q_{5} & =\left(\bar{s}_{L}\gamma_{\mu}\gamma_{\nu}\gamma_{\sigma}b_{L}\right)\underset{q}{\sum}\left(\bar{q}\gamma^{\mu}\gamma^{\nu}\gamma^{\sigma}q\right),\nonumber \\
Q_{6} & =\left(\bar{s}_{L}\gamma_{\mu}\gamma_{\nu}\gamma_{\sigma}T^{a}b_{L}\right)\underset{q}{\sum}\left(\bar{q}\gamma^{\mu}\gamma^{\nu}\gamma^{\sigma}T^{a}q\right),\nonumber \\
Q_{7} & =\frac{e}{16\pi^{2}}m_{b}\left(\bar{s}_{L}\sigma^{\mu\nu}b_{R}\right)F_{\mu\nu},\nonumber \\
Q_{8} & =\frac{g_{s}}{16\pi^{2}}m_{b}\left(\bar{s}_{L}\sigma^{\mu\nu}T^{a}b_{R}\right)G_{\mu\nu}^{a},\nonumber \\
Q_{9} & =\frac{e^{2}}{16\pi^{2}}m_{b}\left(\bar{s}_{L}\gamma^{\mu}b_{L}\right)\underset{l}{\sum}\left(\bar{l}\gamma_{\mu}l\right),\nonumber \\
Q_{10} & =\frac{e^{2}}{16\pi^{2}}m_{b}\left(\bar{s}_{L}\gamma^{\mu}b_{L}\right)\underset{l}{\sum}\left(\bar{l}\gamma_{\mu}\gamma_{5}l\right).\label{eq:1.7.4}
\end{align}

\newpage{}

\section{Soft Collinear Effective Theory (SCET)}

SCET describes the interaction of soft and collinear degrees of freedom
in the presence of hard interactions. The momentum scale for hard
interactions is describe as $Q$. Another important scale called hadronic
scale $\left(\Lambda_{QCD}\right)$ is used for QCD, which is the
scale of hadronization and non-perturbative physics. It is important
to note that $Q\gg\Lambda_{QCD}.$ Soft degrees of freedom have momenta
$p_{soft}^{\mu}$, where $Q\gg\text{\ensuremath{p_{soft}^{\mu}}}.$
If soft modes are non-perturbative then we will have $p\backsim\Lambda_{QCD},$
but if soft modes are perturbative then $\text{\ensuremath{p_{soft}^{\mu}>>\Lambda_{QCD}.}}$
Energetic particles moving in certain preferential direction is derscribed
by collinear degrees of freedom. There are few characteristics of
SCET that make it different from other effective theories in many
aspects. In SCET,
\begin{itemize}
\item The off-shell modes are removed in SCET, but not entire degrees of
freedom.
\item Same particles are defined by multiple fields.
\item Traditionally in EFT, we sum over operators with same power counting
but in SCET, these sums are replaced by integrals, i.e., $\underset{i}{\sum}C_{i}O_{i}=\int d\omega C(\omega)\mathcal{O}(\omega)$. 
\end{itemize}

\subsection{Applications of SCET}

Following are the processes in which SCET can be applied to simplify
the physics:
\begin{itemize}
\item Inclusive and exclusive hard scattering processes.
\item Exclusive jet processes.
\item Charmonium production.
\item inclusive and exclusive $B$-decays.
\end{itemize}

\subsubsection{Light-cone coordinates}

In special relativity, a light cone describes the path that light
would take through spacetime. In light cone coordinate system the
speed of light is equal to 1. Light cone coordinates are also called
null coordinates. In light cone coordinates, the position of an event
in spacetime is represented by two coordinates i.e., the light cone
coordinate, which is also called the time coordinate, and the transverse
coordinate, which shows the position in the plane perpendicular to
the direction of propagation of the light. Let us consider a decay
process i.e., $B\rightarrow D\pi$ in the rest frame of $B$-meson.
This decay takes place by exchanging $W$ boson. In a quark level
process $b\rightarrow c\bar{c}d$, aligning the $\pi$ with the $z$-axis,
its momentum can be expressed aswhere $n^{\mu}=\left(1,0,0,1\right).$
In this case pion has large energy and has 4-momentum close to light
cone. The light cone basis vector i.e., $n$ and $\bar{n}$, must
satisfy the properties following properties 
\[
p_{\pi}^{\mu}=\left(2.310\text{\ensuremath{GeV}},0,0,-2.306\text{ \ensuremath{GeV}}\right)\backsimeq Qn^{\mu},
\]
where $n^{\mu}=\left(1,0,0,1\right).$ In this case pion has large
energy and has 4-momentum close to light cone. The light cone basis
vector $n$ and $\bar{n}$ must satisfy the properties that are
\begin{align*}
n^{2} & =0,\\
\bar{n}^{2} & =0,\\
n.\bar{n} & =2,
\end{align*}
where $\text{\ensuremath{\bar{n}^{\mu}=(1,0,0,-1)}}$. 

Standard 4-vector representation in light cone basis is

\[
p^{\mu}=\frac{n^{\mu}}{2}\bar{n}.p+\frac{\bar{n}^{\mu}}{2}n.p+p_{\perp}^{\mu},
\]
where the $\perp$ components are orthogonal to both $n$ and $\bar{n}.$
Writing

\begin{align*}
p^{\mu} & =\left(p^{+},p^{-},p_{\perp}\right),\\
p^{+} & =p_{+}\equiv n.p,\\
p^{-} & =p_{-}\equiv\ensuremath{\bar{n}}.p.
\end{align*}
The 4-momentum squared is given as

\[
p^{2}=p^{+}p_{-}.
\]

\subsection{Momentum regions in SCET}

In SCET, the momentum regions are the different scales that appear
in a scattering process. In a scattering process there are three different
regions, namely hard, soft and collinear.

\subsubsection{Hard region}

The hard region in SCET, refers to the momenta of particles that are
larger than the momentum scale of the theory. In SCET, such particles
are treated as hard and their interactions can be described perturbatively.

\subsubsection{Soft region}

The soft region in SCET, refers to the momenta of particles that are
smaller than the momentum scale of theory. In SCET, such particles
are treated as soft and effective soft fields are used to describe
their interactions.

\subsubsection{Collinear region }

The collinear region in SCET, refers to the momenta of particles having
collinear modes and effective collinear fields are used to describe
their interactions.

The collinear region in SCET, refers to the momenta of particles having
collinear modes and effective collinear fields are used to describe
their interactions.

Let us put some insight on what the relevant degrees of freedom means
in the context of hard scattering processes. In the rest frame of
the $B$ meson, for $B\rightarrow D\pi$ ,the $B$-meson is composed
of heavy quark, soft quarks and gluons having momenta $\thicksim\Lambda_{QCD}$,
The momentum scaling of pion constituents are 

\[
p^{\mu}\thicksim\left(\Lambda_{QCD},\Lambda_{QCD},\Lambda_{QCD}\right).
\]
By boosting along $-\hat{z}$ component by $\kappa=\frac{Q}{\Lambda_{QCD}}$,
and using light cone coordinates, taking $p^{-}\rightarrow\kappa p^{-}$
and $p^{+}\rightarrow\frac{p^{+}}{\kappa}$, we get

\[
p_{c}^{\mu}=\left(\frac{\Lambda_{QCD}^{2}}{Q},Q,\Lambda_{QCD}\right),
\]
where $,p_{c}^{-}\gg p_{c}^{\bot}\gg p_{c}^{+}$. Whenever the components
of $p_{c}^{\mu}$ obey this, we can say that it scales collinearly
as

\[
p_{c}^{\mu}\thicksim Q\left(\lambda^{2},1,\lambda\right).
\]
For $B\rightarrow D\pi$, we have $\lambda=\frac{\Lambda_{QCD}}{Q}$.
This $\lambda$ will be the power counting parameter of SCET. The
soft momenta of constituents in the $B$ and $D$ meson scale as $p_{s}^{\mu}\thicksim Q(\lambda,\lambda,\lambda).$
Hence, both soft and collinear degrees of freedom are required for
$B\rightarrow D\pi$ decay. 

\subsection{Basic Ingredients for SCET}

Now our aim is to formulate collinear and ultra-soft degrees of freedom
by expanding the full theory i.e., QCD. This will enable us to derive
power counting expressions and check the form of the Lagrangian in
SCET.

\subsubsection{Collinear Spinors}

In physics, a collinear spinor is a type of spinor that is associated
with a particle, moving along a fixed direction. Collinear spinors
are often used in the context of the SCET, which is a framework to
study the dynamics of particles. 

In physics, a collinear spinor is a type of spinor that is associated
with a particle, moving along a fixed direction. Collinear spinors
are often used in the context of the SCET, which is a framework to
study the dynamics of particles. 

We are going to consider the decomposition of Dirac's spinors $u(p)$
for particles and $v(p)$ for antiparticles in collinear limit. For
sake of this, we will derive the collinear spinors via expansion in
momentum components, and subsequently convert the obtained result
into a decomposition of two terms instead of infinite expansion. For
a collinear momentum, we have

\[
p^{\mu}=(p^{0},p^{1},p^{2},p^{3}),
\]
and

\[
p^{-}=p^{0}+p^{3}\gg p^{1,2}\gg p^{+}=p^{0}-p^{3}.
\]
Now the expansion in the momentum components is as follows

\[
\frac{\vec{\sigma}\cdot\vec{p}}{p^{0}}=\sigma^{3}+\cdots.
\]
In above expansion the NLO and NNLO terms are very small, so we can
ignore them and can only rely on leading order term. For particles
the Dirac's spinor is

\begin{align*}
u(p) & =\frac{\left(2p^{0}\right)^{1/2}}{\sqrt{2}}\left(\begin{array}{c}
u\\
\frac{\vec{\sigma}.\vec{p}}{p^{o}}u
\end{array}\right),\\
u_{n} & =\sqrt{\frac{p^{-}}{2}}\left(\begin{array}{c}
u\\
\sigma^{3}u
\end{array}\right).
\end{align*}
Similarly, the Dirac spinor for antiparticles is

\begin{align*}
v(p) & =\frac{\left(2p^{0}\right)^{1/2}}{\sqrt{2}}\left(\begin{array}{c}
\frac{\vec{\sigma}.\vec{p}}{p^{0}}v\\
v
\end{array}\right),\\
v_{n} & =\sqrt{\frac{p^{-}}{2}}\left(\begin{array}{c}
\sigma^{3}v\\
v
\end{array}\right).
\end{align*}
$u$ and $v$ are either $\left(\begin{array}{c}
1\\
0
\end{array}\right)$ or $\left(\begin{array}{c}
0\\
1
\end{array}\right)$. From this, we observed that quark and antiquark remains as relevant
degrees of freedom in the collinear limit both and both spin components
retain their presence in each of the spinors. The contraction of $\gamma$
matrices with $n^{\mu}$ and $\bar{n}^{\mu}$ gives 

\[
\cancel{n}=\left(\begin{array}{cc}
1 & -\sigma^{3}\\
\sigma^{3} & -1
\end{array}\right).
\]
Multiplying with $u_{n}$, we get

\begin{align*}
\cancel{n}u_{n} & =u_{n}\left(\begin{array}{cc}
1 & -\sigma^{3}\\
\sigma^{3} & -1
\end{array}\right),\\
\cancel{n}u_{n} & =0.
\end{align*}
Similarly, if we multiply $\cancel{n}$ with $v_{n}$ we also get
zero. Hence

\begin{align*}
\cancel{n}u_{n} & =0,\\
\cancel{n}v_{n} & =0.
\end{align*}
These can be recognized as the leading term in the equations of motion
$\cancel{p}v(p)=\cancel{p}u(p)=0$, when expanded in the collinear
limit. We define the projection operators that are 

\begin{align*}
P_{n} & =\frac{\cancel{n}\cancel{\bar{n}}}{4},\\
P_{n} & =\frac{1}{2}\left(\begin{array}{cc}
\hat{1} & \sigma^{3}\\
\sigma^{3} & \hat{1}
\end{array}\right).
\end{align*}
Similarly

\begin{align*}
P_{\bar{n}} & =\frac{\bar{\cancel{n}}\cancel{n}}{4},\\
P_{\bar{n}} & =\frac{1}{2}\left(\begin{array}{cc}
\hat{1} & -\sigma^{3}\\
-\sigma^{3} & \hat{1}
\end{array}\right),
\end{align*}
giving

\[
P_{n}u_{n}=\frac{\cancel{n}\bar{\cancel{n}}}{4}u_{n},
\]
where $\frac{\cancel{n}\bar{\cancel{n}}}{4}=1$. Therefore

\[
P_{n}u_{n}=u_{n},
\]
and similarly

\[
P_{n}v_{n}=v_{n}.
\]
The main theme behind this expansion is that when a fermion or an
anti-fermion is produced through hard interactions then there will
be a components that obey spin relations. Now let us decompose the
$\psi$ (Dirac's Fiels) into a field $\xi_{n}$ that obeys the aforementioned
spin relations. So from $\left\{ \gamma^{\mu},\gamma^{\nu}\right\} =2g^{\mu\nu},$
we note that

\[
\frac{\cancel{n}\cancel{\bar{n}}}{4}+\frac{\bar{\cancel{n}}\cancel{n}}{4}=1.
\]
Using this relation we can write $\psi$ in terms of two fields

\begin{align*}
\psi & =P_{n}\psi+P_{\bar{n}}\psi,\\
\psi & =\hat{\xi}_{n}+\phi_{\bar{n}},
\end{align*}
where 

\begin{align*}
\hat{\xi}_{n} & =P_{n}\psi,\\
\hat{\xi}_{n} & =\frac{\cancel{n}\cancel{\bar{n}}}{4}\psi.
\end{align*}
Similarly

\begin{align*}
\phi_{\bar{n}} & =P_{\bar{n}}\psi,\\
\phi_{\bar{n}} & =\frac{\cancel{n}\cancel{\bar{n}}}{4}\psi.
\end{align*}
The following spin relations are satisfied by these fields:
\begin{itemize}
\item $\bar{\cancel{n}}\xi_{n}=0$,
\item $P_{n}\xi_{n}=\xi_{n},$
\item $\cancel{\bar{n}}\phi_{\bar{n}}=0$,
\item $P_{\bar{n}}\phi_{\bar{n}}=\phi_{\bar{n}}.$
\end{itemize}
Here $n$ and $\bar{n}$ refers to collinear directions and $\xi_{n}$
is collinear field in SCET.

\subsubsection{Collinear Fermion Propagator}

The collinear fermion propagator describes the propagation of a fermion
through a spacetime in the collinear limit. The fermion propagator
in the collinear limit is

\[
p^{2}+i_{0}=\bar{n}.pn.p+p_{\perp}^{2}.
\]
 By keeping only large momentum terms and expand the numerator as

\begin{align*}
\text{\ensuremath{\frac{i\cancel{p}}{p^{2}+i_{0}}}} & =\frac{i\cancel{n}}{2}\frac{\bar{n}.p}{p^{2}+i_{0}}+\cdots,\\
\frac{i\cancel{p}}{p^{2}+i_{0}} & =\frac{i\cancel{n}}{2}\frac{1}{n.p+\frac{p_{\perp}^{2}}{n.p}+i_{0}(\bar{n}.p)}+\cdots.
\end{align*}
The coupling of fermion and gluon will be proportional to $\frac{\cancel{\bar{n}}}{2}$.
Therefore, it forms a propagator $P_{n}$ when it combines with the
$\frac{\cancel{n}}{2}$. In above expansion it is seen that both $+i_{0}$
and $-i_{0}$ are present. This indicates that both propagating particles
and antiparticles will be a part of the leading order SCET Lagrangian. 

\subsubsection{Power Counting for Collinear Gluons and Ultra-soft Fields}

Now we are going to analyze the collinear gluon field which is $A_{n}^{\mu}$,
in $n$ collinear basis to find the $\lambda$ scaling of its components.
Covariant gauge gluon propagator in full theory is

\begin{align}
\int d^{4}xe^{ik.x}\left\langle 0\right|TA_{n}^{\mu}(x)A_{n}^{v}(0)\left|0\right\rangle  & =-\frac{i}{k^{2}}\left(g^{\mu v}-\tau\frac{k^{\mu}k^{v}}{k^{2}}\right),\\
 & =\text{\ensuremath{-\frac{i}{k^{4}}}}\left(\text{\ensuremath{k^{2}g^{\mu v}}\ensuremath{-}\ensuremath{\tau k^{\mu}k^{v}}}\right),\label{eq:2.4.2}
\end{align}
where $k^{2}=k_{+}k_{-}+k_{\perp}^{2}=Q^{2}\lambda^{2}$. The $-\frac{1}{k^{4}}$
in Eq. (\ref{eq:2.4.2}) scales as

\begin{align*}
d^{4}x & \thicksim\frac{1}{k^{4}},\\
d^{4}x & \thicksim\frac{1}{\left(k^{2}\right)^{2}}.
\end{align*}
The quantity $\left(\text{\ensuremath{k^{2}g^{\mu v}}\ensuremath{-}\ensuremath{\tau k^{\mu}k^{v}}}\right),$
should be of the same order as the product of $A_{n}^{\mu}(x)A_{n}^{v}(0)$
fields. If both $\mu,v$ indices are perpendicular then $\left(\text{\ensuremath{k^{2}g^{\mu v}}\ensuremath{-}\ensuremath{\tau k^{\mu}k^{v}}}\right)$
is of order $\lambda^{2}$. So we must have $A_{n\perp}^{\mu}\thicksim\lambda.$

\subsubsection{Collinear Wilson Line}

Collinear Wilson line is basically a path-ordered exponential that
describes the gauge transformation of a collinear field as it moves
along a particular path. The collinear Wilson line make gauge-invariant
operators that involve collinear fields, and it is also an important
tool for describing the interactions of collinear particles in SCET.

Consider the process $b\rightarrow ue\bar{\nu}$, in which the $b$
quark is heavy and decays to an energetic collinear $u$ quark. The
weak current associated with this decay is 

\[
J_{QCD}=\bar{u}\Gamma b,
\]
where $\Gamma=\gamma^{\mu}\left(1-\gamma^{5}\right)$.Without gluons
we can match this QCD current onto a leading order current in SCET
by considering the heavy $b$ field to be the $h_{v}$ (HQET field)
and the lighter $u$ field by the $\xi_{n}$ (SCET field). The resulting
SCET operator is

\[
\bar{\xi}_{n}\Gamma h_{v}.
\]
Now, consider the case where an extra $A_{n}^{-}$ gluon is attached
to the heavy quark. The incoming $b$ quark carries momentum $m_{b}v^{\mu}$,
so 
\[
k=m_{b}v+q,
\]
such that
\[
k^{2}-m{}_{b}^{2}=2m_{b}v\cdot q+q^{2}.
\]
Now 

\begin{equation}
\text{\ensuremath{A_{n}^{\mu}}=\text{\ensuremath{\frac{n^{\mu}}{2}}}\ensuremath{\bar{n}}.\ensuremath{A_{n}}+\text{\ensuremath{\frac{\bar{n}^{\mu}}{2}n.}\ensuremath{A_{n}}}+\ensuremath{A}\ensuremath{{}_{\perp}^{\mu}}},\label{eq:2.4.3}
\end{equation}
where in Eq. (\ref{eq:2.4.3}), the $\bar{n}.A_{n}$ , $n.A_{n}$
and $A_{\perp}^{\mu}$ are of order $\lambda^{0},$ $\lambda^{2}$
and $\lambda$, respectively. Including propagator, we can write

\begin{equation}
A_{n}^{\mu A}\bar{\xi}_{n}\Gamma\frac{i(\cancel{k}+m_{b})}{k^{2}-m_{b}^{2}}igT^{A}\gamma_{\mu}h_{\nu}=-g\left(\frac{n^{\mu}}{2}\bar{n}.A_{n}^{A}\right)\bar{\xi}_{n}\Gamma\frac{\left[m_{b}(1+\cancel{v})+\cancel{q}\right]}{2m_{b}v.q+q^{2}}T^{A}\gamma_{\mu}h_{\nu}.\label{eq:2.4.4}
\end{equation}
After expanding Eq. (\ref{eq:2.4.4}), we only consider the lowest
order terms. As $m_{b}v.n\bar{n}.q\thicksim Q^{2}\lambda^{0}$ is
off shell by $\thicksim Q^{2}$, which is hard propagator and will
be excluded while constructing SCET operators. Hence, we can write

\[
A_{n}^{\mu A}\bar{\xi}_{n}\Gamma\frac{i(\cancel{k}+m_{b})}{k^{2}-m_{b}^{2}}igT^{A}\gamma_{\mu}h_{\nu}=-g\bar{n}.A_{n}^{A}\bar{\xi}_{n}\Gamma\left[\frac{m_{b}(1+\cancel{v})+\frac{\cancel{n}}{2}\bar{n}.q}{2m_{b}v.q+q^{2}}+\cdots\right]T^{A}\frac{\cancel{n}}{2}h_{v}.
\]
As $\cancel{n}^{2}=0$ and also $\left(1-\cancel{v}\right)h_{v}=0$,
so

\[
A_{n}^{\mu A}\bar{\xi}_{n}\Gamma\frac{i(\cancel{k}+m_{b})}{k^{2}-m_{b}^{2}}igT^{A}\gamma_{\mu}h_{\nu}=-g\bar{n}.A_{n}^{A}\bar{\xi}_{n}\Gamma\left[\frac{(1-\cancel{v})\frac{\cancel{n}}{2}+v.n}{v.n\cancel{n}.q}+\cdots\right]T^{A}h_{v}.
\]
 Also $\left(1-\cancel{v}\right)h_{v}=0$, giving
\begin{equation}
A_{n}^{\mu A}\bar{\xi}_{n}\Gamma\frac{i(\cancel{k}+m_{b})}{k^{2}-m_{b}^{2}}igT^{A}\gamma_{\mu}h_{\nu}=\bar{\xi}_{n}\left[\frac{-g\bar{n}.A_{n}}{\bar{n}.q}+\cdots\right]\Gamma h_{v}.
\end{equation}
Further mathematical calculations to derive the collinear Wilson line
is not important, so we are going to write it directly as given in
\cite{key-9}
\begin{equation}
W=Pexp\left(ig\stackrel[-\infty]{0}{\int}ds\bar{n}.A_{n}(\bar{n}s)\right).
\end{equation}

\subsection{Symmetries of SCET}

In QFT, symmetries and dimensional analysis plays a vital role in
constructing Lagrangian of a system. The choice of vectors $n_{\mu}$
and $\bar{n}_{\mu}$ is not unique in SCET like in the HQET. By having
these vectors we have a set of transformations for these vectors such
that SCET Lagrangian remains invariant. Both of these transformations
are the reparameterization invariant. For $n_{\mu}$ 

\[
n_{\mu}\rightarrow n_{\mu}+e_{\mu}^{\bot},
\]
where $e_{\mu}^{\bot}$ is of order $\mathcal{O}(\lambda)$ or smaller. 

Writing

\[
n_{\mu}\rightarrow(1+\alpha)n_{\mu},
\]
where $\alpha\thicksim1$. For $\bar{n}_{\mu}$ 
\[
\bar{n}\mu\rightarrow\bar{n}\mu,
\]
\[
\bar{n}_{\mu}\rightarrow(1-\alpha)\bar{n}_{\mu}.
\]
After doing the above exercise we have concluded that if we required
the operators of order $\lambda^{2}$ then, $e_{\mu}\thicksim\lambda$.
While constructing the SCET operators for electroweak theory the importance
of reparameterization invariance can not be ruled out. Reparametrization
symmetry was not obvious in the full theory that is under consideration.
However, while solving the effective theory this symmetry emerges
that is why it is often called emergent symmetry. For the two type
of SCET modes, i.e., soft and collinear, we have apparently two separate
gauge transformations. For soft modes the transformations are \cite{key-10}

\[
U_{s}(x)=exp\left(i\alpha_{s}(x)\right).
\]
Similarly, for collinear modes the transformations are

\[
U_{c}(x)=exp\left(i\alpha_{c}(x)\right).
\]
The scaling of the functions $\alpha(x)$ will depend on the scaling
of the respective fields, so it will be adjust according to the fields
under consideration. The soft fields transformed under soft-gauge
transformation are,

\begin{align*}
q_{s}(x) & =U_{s}(x)q_{s}(x),\\
A_{s}^{\mu}(x) & =U_{s}(x)A_{s}^{\mu}U_{s}^{\dagger}(x)+\frac{i}{g_{s}}U_{s}(x)\left[\partial^{\mu},U_{s}^{\dagger}(x)\right].
\end{align*}
The collinear fields transforms under soft-gauge and these transformation
are

\begin{align*}
\xi(x) & =U_{s}(x)\xi(x),\\
A_{c}^{\mu}(x) & =U_{s}\left(x_{-}\right)A_{c}^{\mu}U_{s}^{\dagger}\left(x_{-}\right).
\end{align*}
Since the fields are invariant under gauge-transformations, it is
necessary to work out the electroweak SCET operators in a gauge-invariant
way. Now moving ahead, the collinear fields will transform under collinear-gauge
transformations as

\begin{align*}
\xi(x) & =U_{c}(x)\xi(x),\\
A_{c}^{\mu}(x) & =U_{c}(x)A_{c}^{\mu}U_{c}^{\dagger}(x)+\frac{i}{g_{s}}U_{c}(x)\left[i\partial^{\mu}+g_{s}\frac{\bar{n}^{\mu}}{2}n.A_{s}\left(x_{-}\right),U_{c}^{\dagger}(x)\right].
\end{align*}
We have transformed the collinear fields under collinear gauge transformations
while the soft fields remains invariant under the these transformations.
It is pertinent to mention that in effective theory two types of symmetries
exist, one is inherent symmetry and other is emergent symmetry. The
symmetry that exists in full theory is called inherent symmetry while
the emergent symmetry is not apparent in the beginning but during
calculations it may appears suddenly as mentioned above.

\subsection{SCET Lagrangian}

We have discussed the dynamics of heavy to light energetic meson decay
in the rest frame of heavy meson with a 4-velocity $v=\left(1,0,0,0\right),$
by using light cone coordinates. In this section, by analyzing and
separating the collinear, ultra-soft gluons and momentum degrees of
freedom, we derive the SCET quark Lagrangian. Let us construct the
zeroth order SCET Lagrangian that should satisfy the following properties:
\begin{itemize}
\item Obtaining the collinear propagator that have proper spin structure.
\item Consist of both collinear quarks and antiquarks.
\item Have interactions with both collinear and ultra-soft gluons.
\item Yielding the correct form of leading order (LO) propagator without
requiring extra expansions.
\end{itemize}
We are going to construct a Lagrangian for the $\hat{\xi}_{n}.$ The
first two requirements that are mentioned above will be satisfied
by it. The standard QCD Lagrangian for massless quarks is \cite{key-11}

\[
\mathcal{L}_{QCD}=\bar{\psi}i\cancel{D}\psi.
\]
Expanding $\psi$ and $D$ in collinear basis yields

\begin{equation}
\mathcal{L}=\left(\bar{\psi}_{n}+\text{\ensuremath{\hat{\xi}}}_{n}\right)\left(\frac{\bar{\cancel{n}}}{2}in.D+\frac{\cancel{n}}{2}i\bar{n}.D+i\cancel{D}_{T}\right)\left(\psi_{\bar{n}}+\hat{\xi}_{n}\right).\label{eq:2.6.1}
\end{equation}
In order to simplify Eq. (\ref{eq:2.6.1}) we are going to use projection
matrix identities for collinear spinors, and by employing this procedure
the various terms vanishes as:

\begin{align*}
\frac{\cancel{n}}{2}i\bar{n}.D\xi_{n} & =0,\\
\bar{\psi}_{\bar{n}}\frac{\cancel{\bar{n}}}{2}in.D & =0.
\end{align*}
Now
\begin{align*}
\text{\ensuremath{\hat{\bar{\xi}}}}_{n}i\cancel{D}_{\perp}\text{\ensuremath{\hat{\xi}}}_{\perp n} & =\text{\ensuremath{\text{\ensuremath{\hat{\bar{\xi}}}}_{n}i\cancel{D}P_{n}\text{\ensuremath{\hat{\xi}}}_{n},}}\\
\text{\ensuremath{\hat{\bar{\xi}}}}_{n}i\cancel{D}_{\perp}\ensuremath{\hat{\xi}}_{\perp n} & =\ensuremath{\hat{\bar{\xi}}_{n}}\ensuremath{iP_{n}\cancel{D}_{\perp}\ensuremath{\hat{\xi}}_{n},}\\
\text{\ensuremath{\hat{\bar{\xi}}}}_{n}i\cancel{D}_{\perp}\text{\ensuremath{\hat{\xi}}}_{\perp n} & =0.
\end{align*}
Similarly
\[
\bar{\psi}_{\bar{n}}i.\cancel{D}_{\perp}\psi=0.
\]
As $\hat{\bar{\xi}}P_{n}=0$ and $\bar{\psi}_{\bar{n}}P_{\bar{n}}=0$,
so from this, the Lagrangian we obtained is

\begin{equation}
\mathcal{L}=\hat{\bar{\xi}}_{n}\frac{\cancel{\bar{n}}}{2}in.D\hat{\xi}_{n}+\bar{\psi}_{\bar{n}}i\cancel{D}_{\perp}\hat{\xi}_{n}+\hat{\bar{\xi}}_{n}i\cancel{D}_{\perp}\psi_{\bar{n}}+\bar{\psi}_{\bar{n}}\frac{\cancel{n}}{2}i\bar{n}.D\psi_{\bar{n}}.
\end{equation}
This Lagrangian is in terms of the $\hat{\xi}_{n}$ and $\phi_{n}$
fields is called QCD Lagrangian. The detail calculations of SCET Lagrangian
are little bit arduous, so for convenience we are going to write the
Lagrangian directly as \cite{key-12}
\[
\mathcal{L}_{n\xi}^{(0)}=e^{-ix.\mathcal{P}}\hat{\bar{\xi}}_{n}\left(in.D+i\cancel{D}_{n\perp}+\frac{1}{i\bar{n}.D_{n}}i\cancel{D}_{n\perp}\right)\frac{\cancel{\bar{n}}}{2}\xi_{n},
\]
where $iD_{n\perp}^{\mu}=\mathcal{P}{}_{\perp}^{\mu}+gA_{n\perp}^{\mu}$
and $i\bar{n}.D_{n}=\mathcal{\bar{P}}+g\bar{n}.A_{n}$ are collinear
covariant derivatives. The SCET Lagrangian, consists of the sum of
terms, each of which represents an interaction or kinetic term for
a particular field. It is pertinent to mention that the form of the
SCET Lagrangian depends on the specific problem under consideration,
but it generally includes terms for the collinear fields, the soft
fields, and the ultra-soft fields. Beside that, the SCET Lagrangian
also includes effective couplings and operators that describe the
interactions between the different field.

\section{Conclusion}

EFT offers a systematic framework for understanding and predicting
the behavior of physical systems across diverse energy scales. Throughout
this article, I have discussed the essential ingredients used to construct
the EFT. Beginning with a exploration of the key concepts, including
the separation of scales, power counting, and the renormalization
group, we have laid the groundwork for a deeper understanding of EFT's
efficacy in describing the dynamics of complex systems. By discussin
the role of effective degrees of freedom and the importance of symmetry
principles, we have highlighted the elegance of the EFT approach.
Moreover, we have briefly discussed the construction of effective
field theory to understand the dynamics of H-atom. Afterwards, I dive
deeply to construct the SCET from QCD. I have discussed momentum regions
of SCET and their particular scalings. Apart from this, the Wilson
line and its significance in the gauge theories is also discussed
in detail. After the whole process of understanding SCET, the general
form of SCET Lagrangian is derived. SCET is used to analyze the decay
amplitudes, branching ratios with great precision.

\end{document}